\documentstyle[preprint,multicol,epsf,aps]{revtex}
\begin{document}
\title{ Specific heat of the ideal gas obeying the generalized 
exclusion statistics}
\author{
Takahiro Aoyama 
\footnote[1]{
E-mail: aoyama@particle.sci.hokudai.ac.jp
\\
Address: Department of Physics,
Hokkaido University,
Sapporo,
060-0810
Japan\\
Tel \& Fax number: +81-11-706-4436}
}
\address{
Department of Physics,
Hokkaido University,
Sapporo,
060-0810
Japan}
\maketitle
\begin{abstract}
We calculate the specific heat of the ideal gas obeying
the generalized exclusion statistics (GES) in the continuum model and 
the tight binding model numerically.
In the continuum model of 3-d space, 
the specific heat increases with statistical parameter at low temperature 
whereas it decreases with statistical parameter at high temperature.
We find that the critical temperature normalized by $\mu_f$ (Fermi energy)
is 0.290. 
The specific heat of 2-d space was known to be independent of $g$ in the
continuum model,
but it varies with $g$ drastically in the tight-binding model.
From its unique behavior, identification of GES particles will be
possible from the specific heat.
\end{abstract}
\newpage
{\flushleft {\bf 1 Introduction}}

The many-particle wave function is symmetric under an exchange of two
identical particles for Boson and is antisymmetric for Fermion.
The dimension of the Hilbert space is determined by counting
the wave functions constrained by the symmetry.
Intermediate statistics are under the intensive study recently.
In one and two-dimensional space,
arbitrary phases can appear in many-particle wave functions 
under an exchange of two identical particles,
which are called anyon \cite{anyon1,anyon2}.
In contrast with the anyon,
Haldane proposed generalized statistics ten years ago 
without specific reference to spatial dimensions \cite{Haldane}.
It is called the generalized exclusion statistics (GES).
In his proposal,
the dimension of the Hilbert space $d$,
and the particle number $N$,
are connected by
\begin{eqnarray}
 \Delta d = - g \Delta N 
\end{eqnarray}
where $\Delta d$ is the change of the available single-particle states number
and $\Delta N$ is the change of the particle number for identical particle
system.
$g$ is a statistical parameter.
We call particles obeying the GES as $g$-on. 
The state-counting of many particle states obeying the GES is proposed later
by Wu, which corresponds to Boson with $g=0$ and Fermion with $g=1$ \cite{Wu}.
This counting rule seems to be exact in the limit 
of a large number of states \cite{Wilczek},
and 
the distribution function of the ideal $g$-on gas
has been derived.

Because the statistics is the fundamental property of nature,
it is important to know if there are particles obeying generalized statistics.
For this purpose,
physical properties of the ideal $g$-on gas should be fully understood.  
Thermodynamic 
properties of the ideal $g$-on gas are those properties and
have been studied by several people.
The mean occupation number was derived in the context of the one-dimensional
solvable model \cite{Isakov a,Isakov b}
and the lowest Landau level anyon model \cite{Das}.
Wu derived the entropy and free energy and so on \cite{Wu},
and Nayak et al. showed the duality \cite{Wilczek}.
The Sommerfeld expansion was applied
\cite{Wilczek,Isakov,iguchi,M. Suzuki}.
Other thermal properties were studied in \cite{Isakov2}.
In particular, the specific heat of the ideal $g$-on gas,
which is the most fundamental observable quantity which may reflect the 
statistics,
has been studied analytically \cite{Wilczek,Isakov,iguchi,M. Suzuki,Isakov2}
in a simple model where the single particle energy is proportional to $p^2$.
The $g$-ons which have more general energy dispersions have not been studied
so far.
In the present work,
we find the specific heat of the ideal $g$-on gas in several spatial 
dimensions of wide energy band system
and narrow energy band system based on numerical calculations.

First,
we study the wide energy band system where the single particle energy is
proportional to {${\bf p}^2$}.
In three-dimensional space, 
we find that the specific heat increases with statistical parameter at low 
temperature region whereas it decreases with statistical parameter at high 
temperature region.
The critical temperature normalized by $\mu_f$ (Fermi energy)
is found to be 0.290,
at which the specific heat is independent of $g$. 
In two-dimensional space, 
the specific heats are independent of the statistical parameter.
The analytical proof of the $g$-independence of the specific heat was given in 
\cite{Isakov,M. Suzuki}.
The origin of the $g$-independence is due to the constant density of state
(DOS) for ideal gas in two-dimensional space.

Next,
we study the ideal $g$-on gas on lattice space model,
which has the narrow energy band,
in two-dimensional space.
The DOS is not constant in this model.
This case would be realized in the fractional quantum Hall state (FQHS).
The quasiparticle appearing in FQHS \cite{Laughlin,Halperin,asw}
is known to be an anyonic soliton with a finite spatial extension. 
Two quasiparticles cannot overlap each other and have a minimum distance.
So it would be reasonable to associate these quasiparticles with the particles
on a lattice and to describe with the tight-binding model.
It was shown, in fact,
that anyons in the lowest Landau level behave as $g$-ons with
a varying DOS \cite{Das}.
Thermodynamic properties of these gas have not been studied.
Thus we study the specific heat of ideal $g$-on gas in  the tight-binding 
model.
By numerical calculations,
we find that the specific heat of this model is totally different from 
the previous cases and varies with a statistical parameter of the GES.

The present paper is organized in the following way.
In section 2,
we review the GES and give the analytical expression
of the distribution functions.
In section 3
numerical calculations of the specific heats of the ideal $g$-on gas
are given in three and two-dimensional space.
In section 4,
the specific heat of the tight-binding model of $g$-on gas is calculated
numerically.
Summary and discussion are given in section 5.
In Appendix A,
we give some useful relations.
In Appendix B,
we give the exact solution for the $g$-on distribution function
of $g=\frac{1}{N}$ where $N$ is a natural number. 
Improvement of the Sommerfeld expansion and the duality of the
coefficient are given in Appendix C.
{\flushleft {\bf 2 Distribution function of $g$-on}}

For Boson and Fermion, 
the number of quantum states of $N$ identical particles occupying
$G$ states is given by 
\begin{eqnarray}
 W_b &=& \frac{(G+N-1)!}{N!(G-1)!}  ,\\
 W_f &=& \frac{G!}{N!(G-N)!}
\end{eqnarray}
respectively. 
Wu proposed \cite{Wu} the interpolation formula implying the GES as
\begin{eqnarray}
 W_g = \frac{[ G+(N-1)(1-g) ]!}{N! [ G- gN -(1-g) ]!}
  \label{eqn:countstate}
\end{eqnarray}
with $g=0$ corresponding to Boson and $g=1$ Fermion. 
This is the state-counting rule for $g$-on.

Using Eq. (\ref{eqn:countstate}),
the distribution function reads
\begin{eqnarray}
 f(\epsilon) &=& \frac{1}{ y(e^{\beta(\epsilon - \mu)})+ g}
   \label{eqn:diseq}
\end{eqnarray}
where the function $y$ satisfies the functional equation
\begin{eqnarray}
 y( \zeta )^g (\: 1+y( \zeta )\: )^{1-g}
 =
 \zeta,
  \quad \zeta=e^{\beta(\epsilon-\mu)} . 
 \label{eqn:y}
\end{eqnarray}
From Eq. (\ref{eqn:y}) $y( \zeta )= \zeta$ for $g=1$,
and $y( \zeta )=\zeta-1$ for $g=0$,
respectively.
We can easily obtain other exact solutions of Eq. (\ref{eqn:y}) for
$g=2, 3, 4, \frac{1}{2}, \frac{1}{3}, \frac{1}{4}$
by elementary calculation. 
Moreover,
the other solutions for
$g=\frac{1}{N}$ where $N$ is an arbitrary
natural number ($N>1$) are given by 
\begin{eqnarray}
 \bar{y}
&=&
\left( 1-\frac{1}{N} \right)
\left[
_{N -1} F _{N-2}({\scriptstyle \frac{1}{N},\frac{1}{N},\frac{2}{N},\cdots
\frac{N-2}{N}
;
\frac{1}{N-1},\frac{2}{N-1},\cdots
\frac{N-2}{N-1}; \frac{-N^{\scriptscriptstyle N} 
 e^{\scriptscriptstyle N(x-p)}}
{(N-1)^{\scriptscriptstyle N-1}}
})
-1\right]
\end{eqnarray}
where 
$x\equiv \frac{\epsilon}{t}$, $p\equiv \frac{\mu}{t}$ and
$_l F_n (\alpha_1,\cdots,\alpha_l , \beta_1,\cdots,\beta_n;z)$
is the hypergeometric function.
The temperature $t$ is normalized by $\mu_f$ ($\beta\mu_f \equiv \frac{1}{t}$).
The derivation of $\bar{y}$ is given in Appendix B.

The positivity of $\zeta$ leads to the generalized exclusion principle
\begin{eqnarray}
 f(\epsilon) \leq \frac{1}{g} . 
\end{eqnarray}
At zero temperature, 
we get $\zeta=0$ and $y(\zeta)=0$ for $\epsilon-\mu_g \leq 0$, 
and $\zeta=\infty$ and $y(\zeta)=\infty$ for $\epsilon-\mu_g \geq 0$, 
where $\mu_g$ is a pseudo-Fermi energy for arbitrary $g$ cases. 
Then the value of the average occupation number
at zero temperature is
\begin{eqnarray}
 f(\epsilon) = \left\{
\begin{array}{ll}
  \frac{1}{g} &  (\epsilon-\mu_g \leq 0),\\
  0 & (\epsilon-\mu_g \geq 0).
\end{array}
\right. 
 \label{eqn:gep}
\end{eqnarray}
Eq. (\ref{eqn:gep}) means that one particle occupies $\frac{1}{g}$ states 
at zero temperature. 
{\flushleft{
\bf 3 The specific heat of the ideal $g$-on gas in 3-d and 2-d space}}

We numerically compute the specific heat of the ideal $g$-on gas 
in 3-$d$ and 2-$d$ space in this section.
We assume that the particle of any GES has the same mass
and 
the spectrum, $\frac{p^2}{2m}$,
and neglect the spin degree of freedom. 
In $d$-dimensional space,
the DOS for the system with particle number, $N$, is
$ D(\epsilon)= q(d) \epsilon^{\frac{d}{2}-1}$
where $q(d)=\frac{Nd}{2} \mu_f^{-\frac{d}{2}}$
and $\mu_f$ is Fermi energy given by 
Eq. (\ref{eqn:fermienergy}) in Appendix A.
The chemical potential $\mu(T)$ is determined by the particle number
$N =  \int_0 ^\infty D(\epsilon) f(\beta(\epsilon - \mu)) d\epsilon$.
Using the $\mu(T)$,
we can calculate the average 
energy $E$ and the specific heat $C$ from next relations
$E =  \int_0 ^\infty D(\epsilon) \;\epsilon \;
                          f(\beta(\epsilon - \mu)) d\epsilon$
and
$ C = -k_B \beta^2 \frac{dE}{d\beta}$
where $\beta=\frac{1}{k_B T}$. 
Converting the integral variables $\epsilon$ to $y$ for $N$ and $E$
gives the next relations
\begin{eqnarray}
 \frac{2}{d} = 
t^{\frac{d}{2}} \int^\infty _a dy
     \frac{ 
      \left\{
        \log{( \frac{y}{a} )^g ( \frac{y+1}{a+1} )^{1-g} } 
      \right\}^{\frac{d}{2}-1}
          }
         {
          y(y+1)
         } 
  \label{eqn:aa}
\end{eqnarray}
and
\begin{eqnarray}
 \frac{E}{N \mu_{f} } = \frac{d}{2} t^{\frac{d}{2}+1}
   \int^\infty _a dy
     \frac{ 
           \left\{ \log{( \frac{y}{a} )^g ( \frac{y+1}{a+1} )^{1-g}}
           \right\}^{\frac{d}{2}}
          }
         {
          y(y+1)
         } . 
   \label{eqn:ene}
\end{eqnarray}
$a$ corresponds to the value of $y$ for $\epsilon=0$, 
which satisfies 
$ -\beta \mu = g \log{a} + (1-g) \log{(a+1)} $.
For arbitrary $g$-on except Boson, 
we can solve Eq. (\ref{eqn:aa}) with respect to $a$ at fixed temperature $t$
numerically and obtain the average energy from Eq. (\ref{eqn:ene}).

{\flushleft{$d=3$}}

\begin{figure}
\vspace{-3cm}
\centerline{
\epsfxsize=6in \epsffile{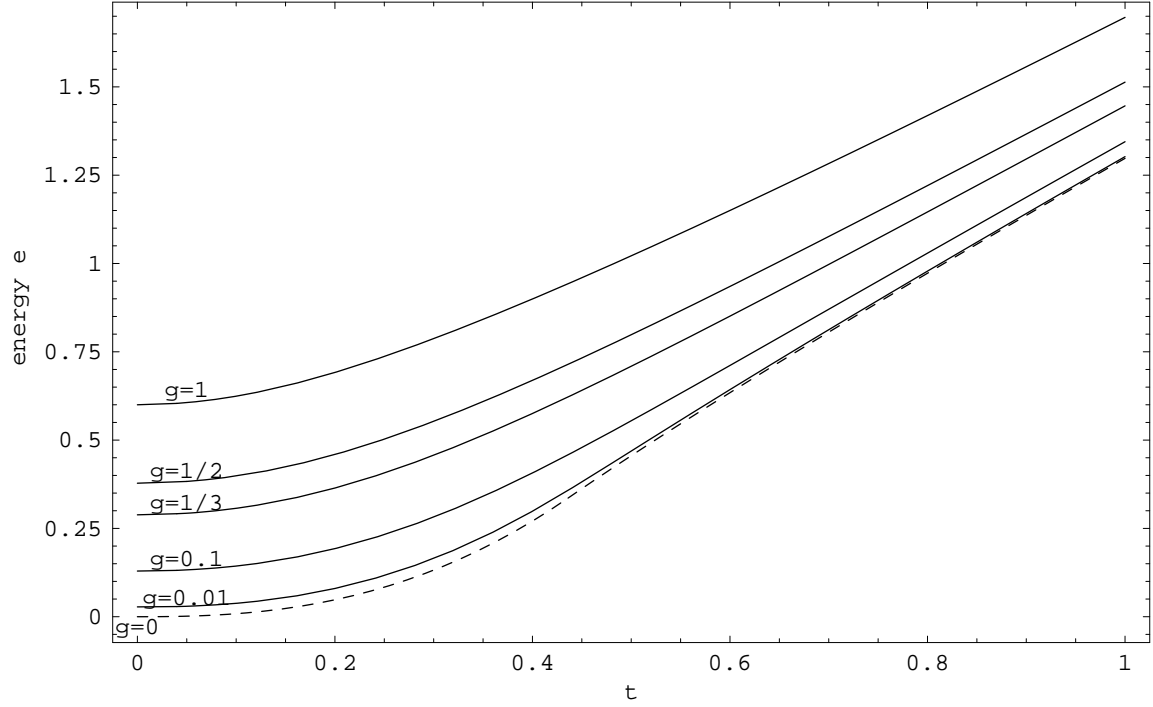}}
\vspace{-3cm}
\caption{Temperature dependence of energy in 3-d space.
The vertical axis is $e\equiv \frac{E(t)}{N\mu_f}$ and 
the horizontal axis is temperature normalized by $\mu_f$.
The broken line represents the Boson case ($g=0$).}
\label{fig:3dene}
\end{figure}

For three-dimensional space,
the average energy $E(t)$ is shown in Fig. \ref{fig:3dene}.
At low temperature, 
the average energy increases with $g$,
which reflects the generalized exclusion principle. 
At high temperature, 
all curves have the same slope and
they go to the classical limit. 

\begin{figure}
\vspace{-3cm}
\centerline{
\epsfxsize=6in \epsffile{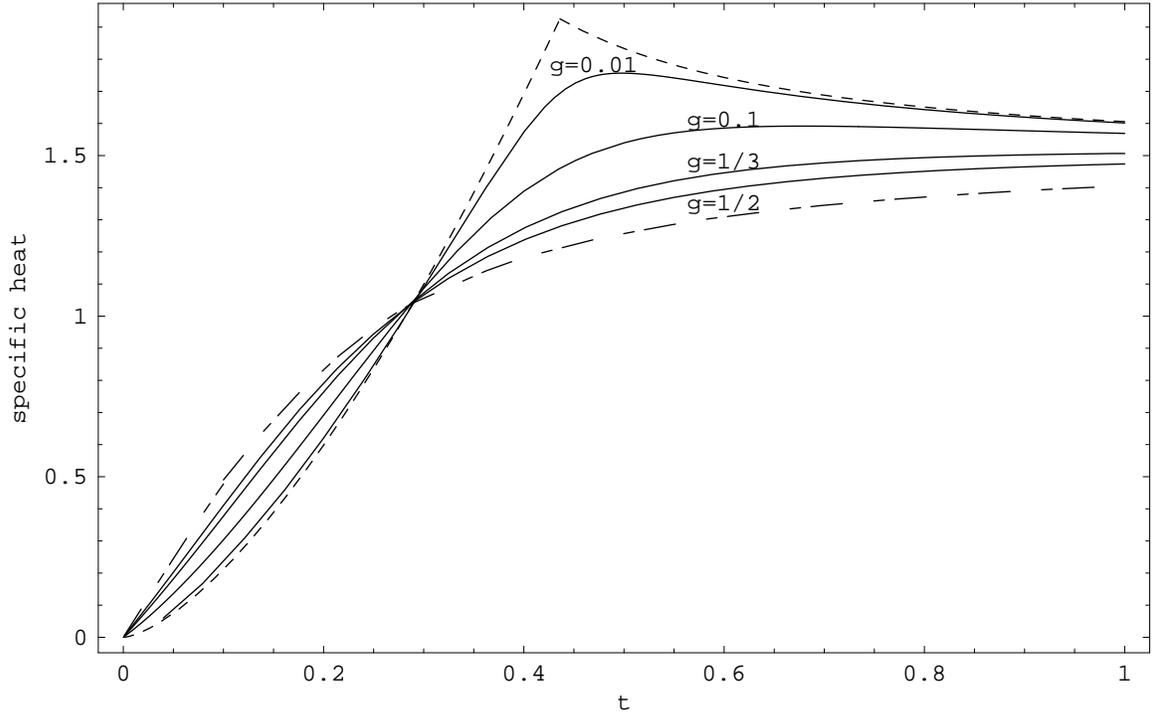}}
\vspace{-3cm}
\caption{Temperature dependence of specific heat in 3-d space.
The vertical axis is $c\equiv \frac{C}{k_B N}$ and
the horizontal axis is temperature normalized by $\mu_f$.
The dashed line represents the Boson case, 
and the dashed and dotted line represents the Fermion case. 
The critical temperature is 0.290.
}
\label{fig:3spe}
\end{figure}

The specific heat versus temperature is shown in Fig. \ref{fig:3spe}. 
The specific heat changes continuously
and has one fixed point at $t=0.290$.
The entropy increases with $g$ below this temperature,
but decreases with $g$ above this temperature.
This reflects the generalized exclusion principle. 
For $g=0$ case (Boson), 
infinite particle numbers in the ground state cause the Bose-Einstein 
condensation. 
For $g \neq 0$ case, 
no condensation occurs at low temperature as mentioned in \cite{iguchi}.
The behavior of the specific heat is different for different statistics,
so in principle we are able to identify the GES from the 
specific heat. 

\begin{figure}
\vspace{-3cm}
\centerline{
\epsfxsize=6in \epsffile{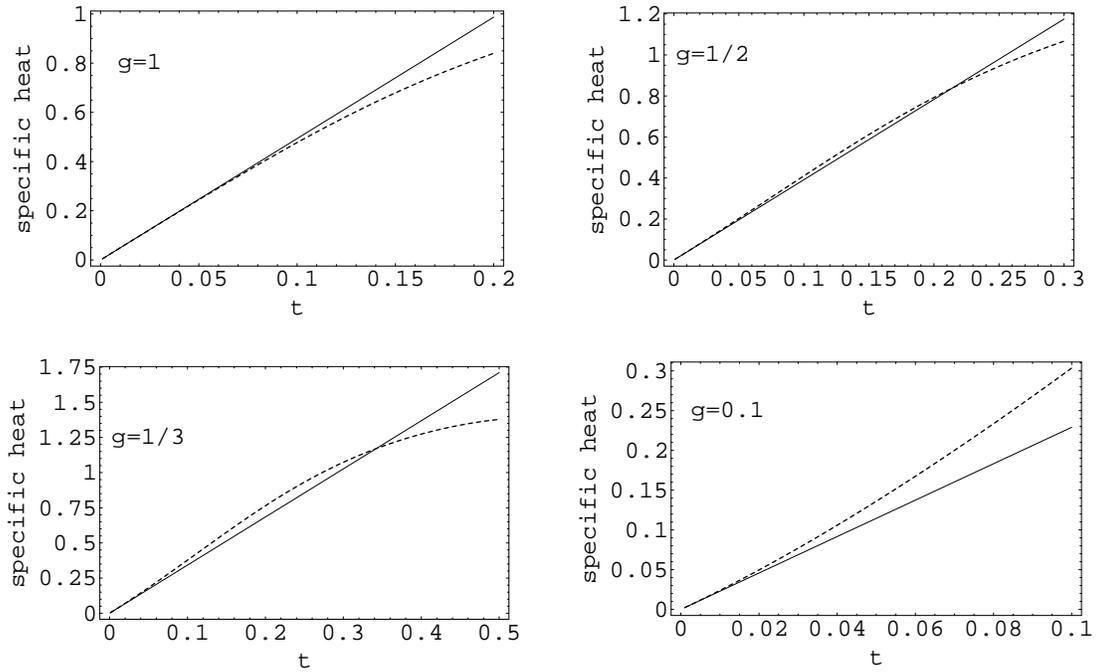}}
\vspace{-3cm}
\caption{Comparison of the Sommerfeld expansion with the numerical calculation
in 3-d space.
The first term in the expansion is written by a solid line and
numerical results are given by a dotted line.
The deviation is large for small $g$.
The first term in the expansion is $\frac{C}{k_B N}=\frac{\pi^2}{2}g^{1/3} t$.
Note that the scales of $t$ are different for each figures.}
\label{fig:som1}
\end{figure}

Some arguments \cite{Wilczek,Isakov,iguchi,M. Suzuki},
in which the specific heat of the ideal $g$-on gas is expressed as a power
series of the temperature
(Sommerfeld expansion for Fermion),
suggest that the coefficient of the first power of the temperature
increases with $g$ in 3-d space
(we improve the Sommerfeld expansion and show the duality of the coefficient
in Appendix C).
Our calculation is not based on the Sommerfeld expansion,
and it is possible to check the validity of the Sommerfeld expansion.
The deviation between the linear part in the temperature and 
numerical result is given in Fig. \ref{fig:som1}.
At low temperature,
the deviation is small,
so the Sommerfeld expansion is good.
However,
the deviation decreases with $g$.
For the Boson's case below the Boson's critical temperature it behaves as
$t^{3/2}$.
So it may be natural to see that the specific heat for very small $g$ 
differs from the linear temperature dependence.
Hence the Sommerfeld expansion is not good for small $g$.


{\flushleft{$d=2$}}

\begin{figure}
\vspace{-3cm}
\centerline{
\epsfxsize=6in \epsffile{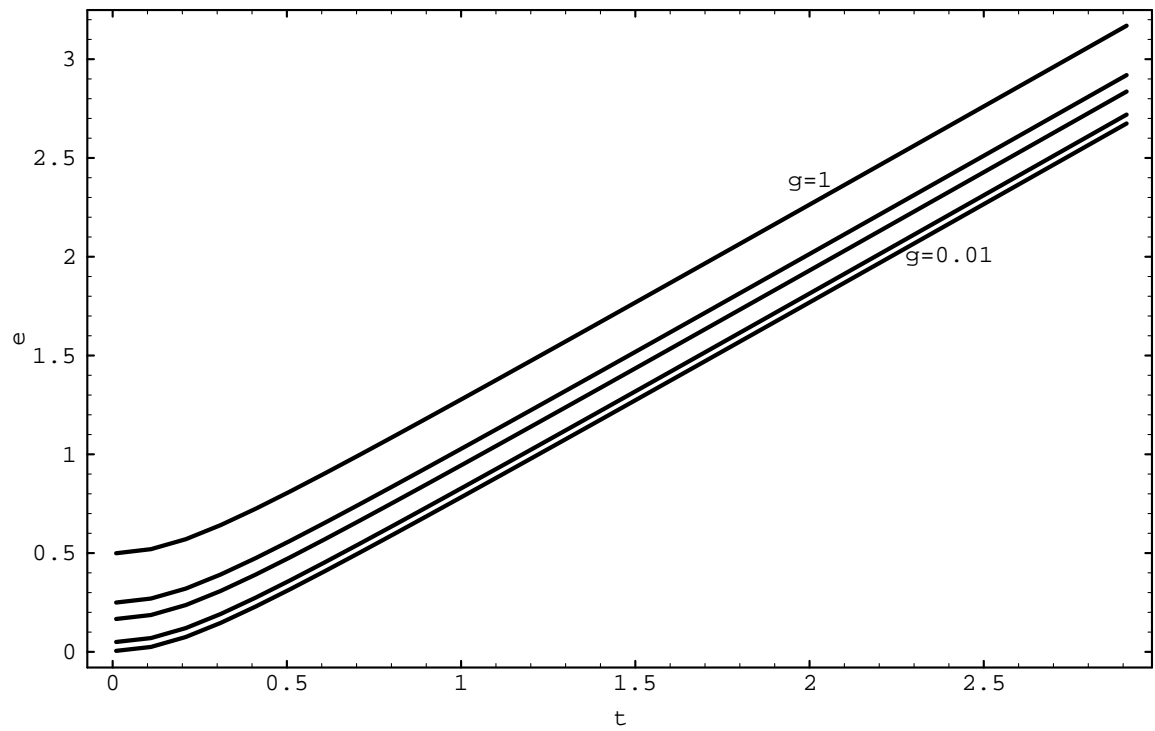}}
\vspace{-3cm}
\caption{Temperature dependence of energy in 2-d space.
All lines are parallel.
They differ by constant.
From top to bottom, 
$g=1, \frac{1}{2}, \frac{1}{3}, 0.1, 0.01$.}
\label{fig:2denergy}
\end{figure}

\begin{figure}
\vspace{-3cm}
\centerline{
\epsfxsize=6in \epsffile{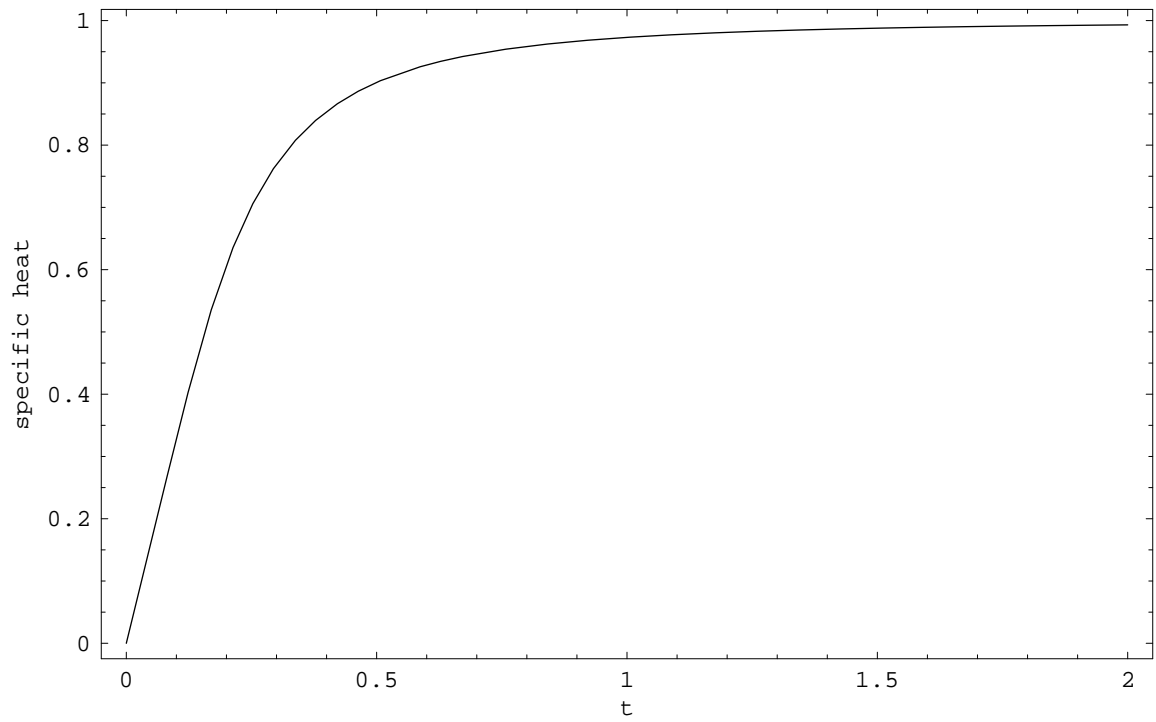} }
\vspace{-3cm}
\caption{Temperature dependence of the specific heat in 2-d space
($g=1, \frac{1}{2}, \frac{1}{3}, 0.1, 0.01$).
All lines overlap.}
\label{fig:2dspe}
\end{figure}

For two-dimensional space,
the average energy $E(t)$ is plotted in Fig. \ref{fig:2denergy}.
At zero temperature, 
$E$ reflects the generalized exclusion principle and depends on $g$. 
At high temperature it approaches to the classical limit. 
The specific heat is shown in Fig. \ref{fig:2dspe}.
Figure \ref{fig:2dspe} shows that
the specific heat is independent of the statistical parameter $g$
at arbitrary temperature.
The analytical proof of the $g$-independence of the specific heat
has been given in \cite{Isakov,M. Suzuki},
using the $\epsilon$-independence of the DOS
($ D(\epsilon) \nonumber = \frac{N}{\mu_{f}}$ for 2-$d$ space).

{\flushleft{\bf 4
Specific heat of the tight-binding model of the ideal $g$-on gas
in 2-d space}}

The quasiparticle of Laughlin's theory
of the FQHS has a finite energy width and is 
approximately described by the lattice model. 
Girvin et al.\cite{Girvin} calculated the energy spectrum
of the quasiparticle of the Laughlin's theory numerically,
based on the single mode approximation analogous to the Feynman's theory of superfluid helium.
Their result implies that the energy spectrum of the excited state in
FQHS seems to be explained by the
tight-binding model.
Laughlin's theory describes the FQHS
($\nu$ means the filling factor of the FQHS)
of $\nu = \frac{1}{3}$ beautifully.
In this case, 
the quasiparticle has $\frac{e}{3}$ charge \cite{Laughlin,asw}, 
and satisfies fractional statistics of $\frac{1}{3}$.
Hence there may be three quasiparticles 
in one available single-particle state;
that is $g=\frac{1}{3}$ statistics.
It would be reasonable to express
the quasiparticle in the Laughlin's theory by the tight-binding of $g$-on gas.
We study the $g$-on in the tight-binding model,
consequently.

In the tight-binding model,
the energy spectrum is 
${\scriptstyle \epsilon =  -c ( \cos(k_x b)+\cos(k_y b) )}$
and 
the DOS of this model is given by
\begin{eqnarray}
 D(\epsilon)=\frac{M}{c\pi^2}\; K \! \! \left( 
 \frac{1}{2}\sqrt{4-\left(
    \frac{\epsilon}{c} \right)^2} \right)
\end{eqnarray}
where $c$ is the hopping constant and 
$b$ is the lattice spacing,
and $M(\equiv \frac{S}{b^2})$ is the number of the lattice.
$K(x)$ is the complete elliptic integral of the first kind
\begin{eqnarray}
 K(x) &=& \int^{\frac{\pi}{2}} _0 d\phi \frac{1}{\sqrt{1-x^2 \sin^2 \phi}}
.
\end{eqnarray}
The particle number and the average energy of this model is
given by
\begin{eqnarray}
N
&=&
 \int_{-2c} ^{2c} d\epsilon
 D(\epsilon) f(\epsilon) \nonumber
=
\frac{M}{c\pi^2} \int_{-2c} ^{2c} d\epsilon \;
 K \! \! \left( \sqrt{1- \left(\frac{1}{2} 
 \frac{\epsilon}{c} \right)^2} \right)
\frac{1}{y(\zeta)+g},
\\
 E 
&=&
 \int_{-2c} ^{2c} d\epsilon
 D(\epsilon) \epsilon f(\epsilon) \nonumber
=
\frac{M}{c\pi^2} \int_{-2c} ^{2c} d\epsilon \;
 K \! \! \left( \sqrt{1-
 \left( \frac{1}{2} \frac{\epsilon}{c} \right)^2} \right)
\frac{\epsilon}{y(\zeta)+g}.
\end{eqnarray}
It is very difficult to calculate the specific heat for arbitrary value
of $g$.
Hence we select special values of $g$,
$g=1,\frac{1}{2},\frac{1}{3},\frac{1}{4}$.
We fix $\frac{N}{M}=\frac{1}{2}$,
which corresponds to the half-filling 
state for Fermion and the chemical potential of Fermion is zero. 
We numerically calculate
the temperature dependence of the average energy by using $\mu(t)$
and obtain the specific heat. 

\begin{figure}
\vspace{-3cm}
\centerline{
\epsfxsize=6in \epsffile{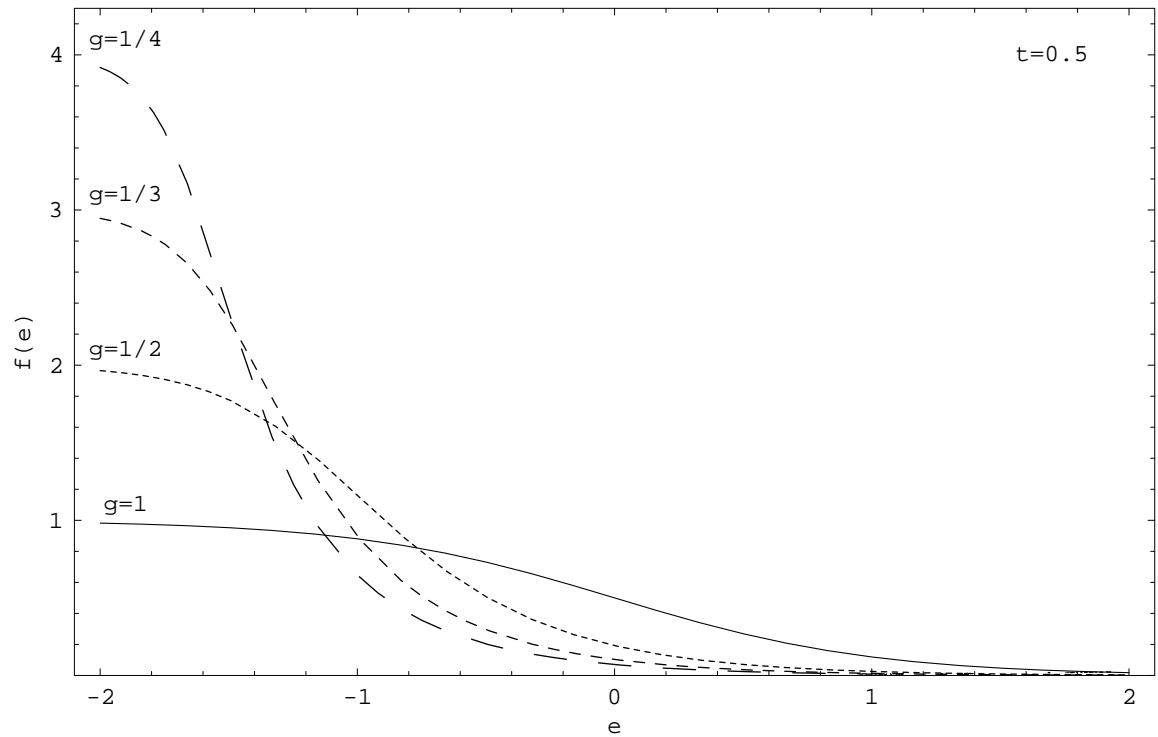}}
\vspace{-3cm}
\caption{The distribution function at $t=0.5$. 
The vertical axis is the distribution function $f(\epsilon)$ and
the horizontal axis is one-particle energy normalized by the hopping constant
($e \equiv \frac{\epsilon}{c}$).}
\label{fig:bunpu}
\end{figure}

\begin{figure}
\vspace{-3cm}
\centerline{
\epsfxsize=6in \epsffile{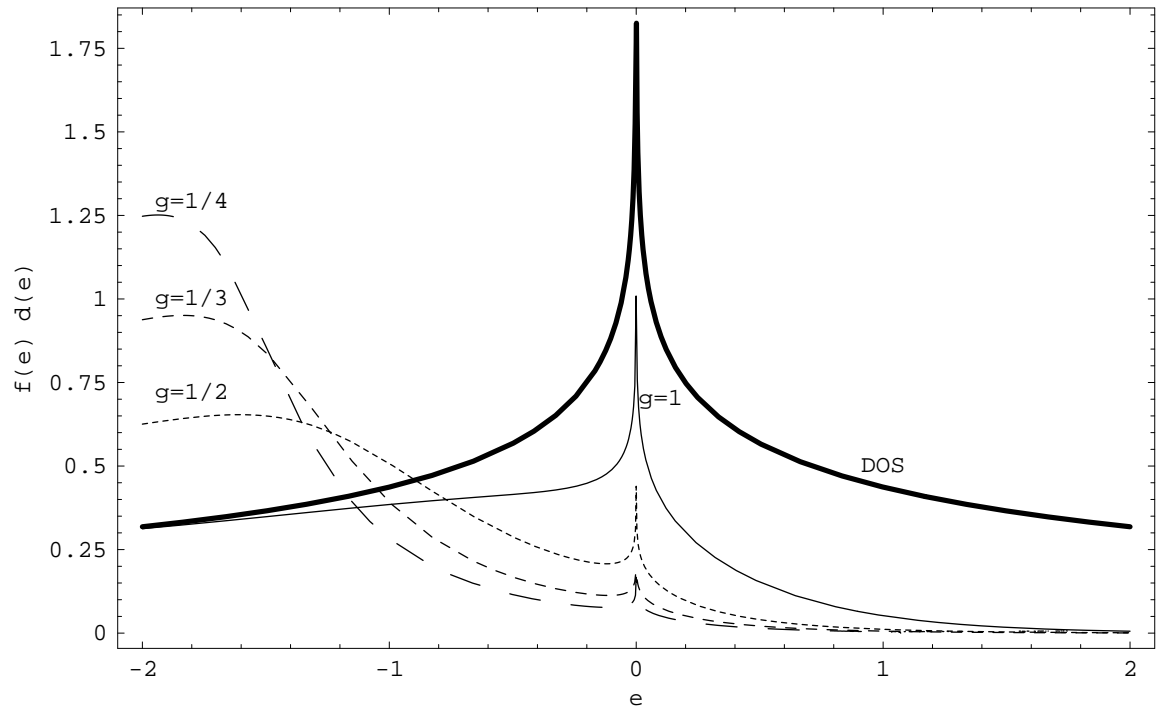}}
\vspace{-3cm}
\caption{The product of $f(\epsilon)$ and $d(\epsilon)$ at $t=0.5$.
The vertical axis is the product $f(\epsilon)$ and 
$d(\epsilon) \equiv \frac{D(\epsilon)}{N}$ and
the horizontal axis is one-particle energy normalized by the hopping constant.}
\label{fig:fd}
\end{figure}

\begin{figure}
\vspace{-5cm}
\centerline{
\epsfxsize=5in \epsffile{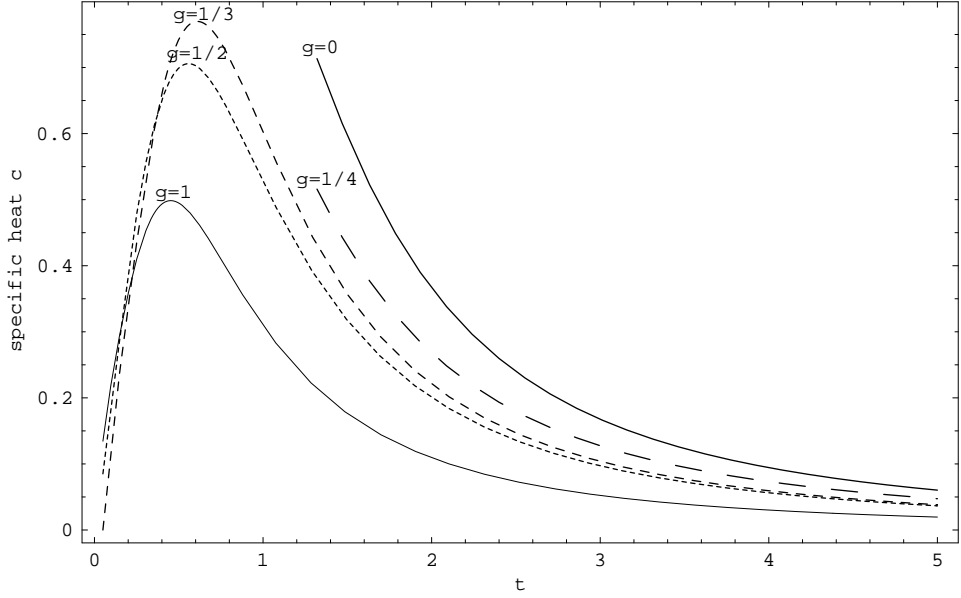}}
\vspace{-4cm}
\caption{The specific heat of the tight-binding model of the ideal $g$-on gas.
The vertical axis is $c\equiv \frac{C}{k_B N}$ and
the horizontal axis is temperature normalized by the hopping constant.}
\label{fig:tight}
\end{figure}

\begin{figure}
\centerline{
\epsfxsize=3in \epsffile{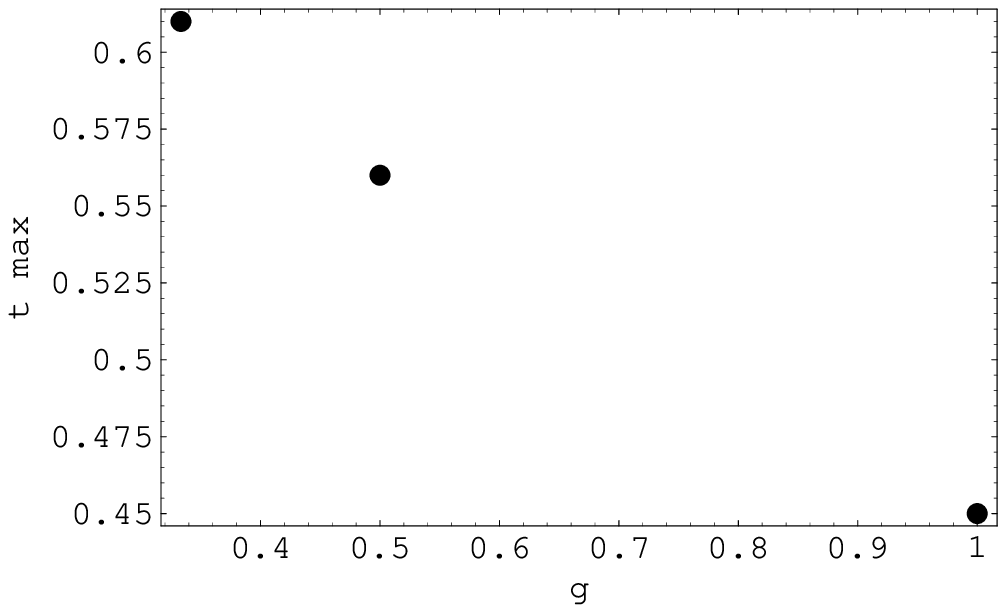}
\hspace{1cm}
\epsfxsize=3in \epsffile{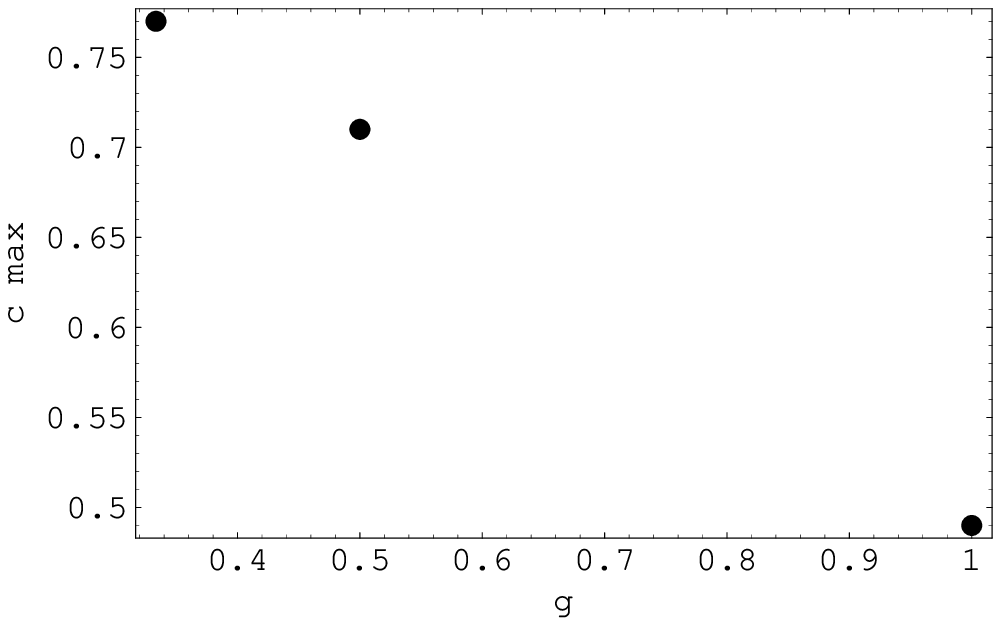}}
\caption{The temperature where the specific heat becomes maximum and
its value.}
\label{fig:max}
\end{figure}

The result is given in Fig. \ref{fig:tight}.
We cannot obtain the specific heat of $g=\frac{1}{4},0$ at low temperature
range now owing to technical problem of numerical calculation.
In the high temperature limit,
the one-particle energy goes to zero.
In this limit,
the distribution of the low energy particle number is very small,
because many particles have much of energy.
The integral decreases at high temperature,
since the integral region is restricted between $-2c$ and $2c$. 
As a result,
the specific heat vanishes in the low temperature limit.
On the other hand,
the specific heat becomes maximum at one temperature value
of order $1$; 
that is, 
when the temperature goes to the order of the 
hopping constant
(the temperature is normalized by the hopping coupling $c$), 
the fluctuation of energy is very large.
From the DOS in Fig. \ref{fig:fd}, 
we read that the most of the 
contribution comes from the particle having the energy of the order of
hopping constant.
This explains why $C(t)$ has a maximum qualitatively.
$C(t)$ becomes maximum at $t=0.45, 0.56, 0.61$ for $g=1, \frac{1}{2}, 
\frac{1}{3}$.
The maximum values are 0.49, 0.71, 0.77 (Fig. \ref{fig:max}).
The difference of the maximum values
reflect each statistics.
See Fig. \ref{fig:bunpu}.

From our results,
$g=\frac{1}{3}$ case,
which will be realized in the $\nu=\frac{1}{3}$ FQHS,
is different from Fermion case.
Therefore,
we will be able to observe the exotic statistics $g=\frac{1}{3}$
by measuring the specific heat of the $\nu=\frac{1}{3}$ FQHS
at the temperature for the order of the hopping constant.
However,
it would be difficult to observe
the specific heat of 2-d electron system,
since the thermal effect of the orthogonal directions
to the plane affects the plane's thermal phenomena, 
in which the FQHS is realized.
{\flushleft{\bf 5 Summary and Discussion}}

The continuum model in three-dimensional space shows that
the specific heat depends on statistics.
Therefore,
we will be able to observe the signature of
exotic statistics by measuring the specific heat.
Moreover,
we find that there is a critical 
temperature $t=0.290$ by numerical calculation.
Above this temperature the specific heat 
increases with the statistical parameter but below this temperature
the specific heat decreases.

As was shown before,
the continuum model in  two-dimensional space shows that 
the specific heat does not depend on statistics.
The $g$-independence is caused by the constant DOS.
In the tight-binding model, 
where the DOS is not constant,
the specific heat of the ideal  $g$-on gas depends
on statistics even in two-dimensional space.
Hence we are able to distinguish the GES by measuring the value of the specific
heat especially from the peak of the specific heat.
The temperature and the value of the specific heat in which the specific heat 
becomes maximum is obtained for several $g$ values. 
It would be exciting to identify the exotic statistics.

{\small T.-H. A. is grateful to K. Ishikawa for suggesting these subjects
and careful reading of this manuscript 
and
also thank N. Maeda and J. Goryo for inspiring and critical discussions.
It is a pleasure to thank H. Suzuki for very useful discussions.}
\appendix
{\flushleft {\bf Appendix A: Some useful relations}}

From Eq. (\ref{eqn:y}) $y$ satisfies 
\begin{eqnarray}
 y(\zeta)^g (\; 1+y(\zeta) \;)^{1-g}
 = \zeta
\end{eqnarray}
where $\zeta = e^{\beta(\epsilon-\mu)}$.
Taking logarithm of it gives
\begin{eqnarray}
 \beta(\epsilon-\mu)= g \log{y} + (1-g)\log{(y+1)} .
           \label{eqn:e-y}
\end{eqnarray}
We convert the integral variable $\epsilon$ into $y$.
From Eq. (\ref{eqn:e-y})
we obtain 
\begin{eqnarray}
 d\epsilon = \frac{1}{\beta} \frac{y+g}{y(y+1)} dy
 . 
  \label{eqn:ey}
\end{eqnarray}

The density of state $D(\epsilon)$ of the ideal gas in $d$-dimensional
space is given by
$
 D(\epsilon) = \frac{(2m \pi)^{\frac{d}{2}}}{\Gamma(\frac{d}{2})}
  \left(\frac{L}{h}\right)^d  \epsilon^{\frac{d}{2}-1} 
$
where $h$ is the Planck constant.
The Fermi energy is determined by 
$
   N = \int^{\mu_{f}} _0 d\epsilon D(\epsilon)
$
is
\begin{eqnarray}
 \mu_f = \left\{
      \Gamma\left(\frac{d}{2}+1\right) \rho
         \right\}^{\frac{2}{d}} \frac{h^2}{2m\pi},
 \label{eqn:fermienergy}
\end{eqnarray}
where $\rho=\frac{N}{V}$ ($V=L^d$).
Especially,
in three and two-dimensional space,
the Fermi energy is given by 
\begin{eqnarray}
  \mu_f &=& \left( \frac{3\rho}{4 \pi} \right)^{\frac{2}{3}}
            \frac{h^2}{2m},
  \label{eqn:fermienergy3d}
        \\
  \mu_f &=& \frac{\rho}{\pi} \frac{h^2}{2m}.
  \label{eqn:fermienergy2d}
\end{eqnarray}
{\flushleft {\bf Appendix B: 
Exact solution of the distribution function for $g=\frac{1}{N}$}}

We obtain the analytical solution of $y$ in this Appendix \cite{H. Suzuki}.
In particular,
in the case of $g=\frac{1}{N}$ where $N$ is a natural number (except $N=1$),
$y$ is represented by the hypergeometric function.

In general,
the $i$-th solution $\bar{z_i}$ of $f(z)=0$,
where $z$ is a complex variables,
is gained by
\begin{eqnarray*}
  \bar{z_i}=\frac{1}{2\pi i} \oint dz 
  \frac{ f^{\prime} (z)}{f(z)} z 
\end{eqnarray*}
where the contour of the integral is circle around $z_i$.
From Eq. (\ref{eqn:y})
in low temperature limit and $\epsilon -\mu \leq 0$,
we see that $e^{\frac{\epsilon-\mu}{t}} \rightarrow 0$ and
then $y \simeq e^{\frac{\epsilon-\mu}{t}}$.
So the solution of $y$ in low temperature limit
will behave as $e^{\frac{\epsilon-\mu}{t}}$ and go to zero.
Using this formula and analytical continuation of $y$,
the solution is given by
\begin{eqnarray}
  \bar{y} = \frac{1}{2\pi i} \oint dy \quad
 \frac{(y+1)^{\frac{1}{g}-1}+(\frac{1}{g}-1) \quad y (y+1)^{\frac{1}{g}-2}}
{y (y+1)^{\frac{1}{g}-1}-e^{\frac{x-p}{g} }} \quad y 
 \label{eqn:ysolution}
\end{eqnarray}
where the contour of the integral is circle around $0$ and
$x\equiv \frac{\epsilon}{t}$, $p\equiv \frac{\mu}{t}$.
In low temperature limit,
Eq. (\ref{eqn:ysolution}) is expanded by 
$\frac{y}{e^{x-p}}$.
The residue gives the integral value.
After short calculation,
the solution $\bar{y}$ is given by 
\begin{eqnarray}
  \bar{y}
&=&
 \sum_{n=1} ^{\infty}
\frac{ \Gamma({\scriptstyle -\frac{n}{g}+n+1} ) }{ 
\Gamma({\scriptstyle -\frac{n}{g}+2 }) }
\frac{( e^{\frac{x-p}{g} } )^n }{n!}
.
\label{eqn:ysolution2}
\end{eqnarray}
Because the singularity of the solution depends on $g$,
we consider the case of $g=\frac{1}{N}$ where $N$ is natural number
to avoid the singular expression of Eq. (\ref{eqn:ysolution2}).
By using the important relation of Gamma function
$
  \Gamma({\scriptstyle z})\Gamma({\scriptstyle 1-z}) = \frac{\pi}{\sin{\pi z}}
$,
the solution is represented by a nonsingular form.
The procedure of the calculation is as follows;
\begin{eqnarray*}
  \bar{y}
&=&
 \sum_{n=1} ^{\infty}
\frac{ \Gamma({\scriptstyle -\frac{n}{g}+n+1 }) }{ 
      \Gamma({\scriptstyle -\frac{n}{g}+2} ) }
\frac{( e^{\frac{(x-p)}{g} } )^n }{n!}
\\
&=&
- \sum_{n=1} ^{\infty}
\frac{ \Gamma({\scriptstyle nN-1} ) }{ \Gamma({\scriptstyle nN-n} ) }
\frac{(- e^{N(x-p)} )^n }{n!}
\\
&=&
- \sum_{n=1} ^{\infty}
\frac{(- e^{N(x-p)} )^n }{n!}
\frac{
  \frac{N^{N \left(n-\frac{1}{N} \right)} }
       { (2 \pi)^{\frac{N-1}{2}} \sqrt{N} }
  \Gamma({\scriptstyle n-\frac{1}{N}} )\Gamma({\scriptstyle n} )
  \Gamma({\scriptstyle n-\frac{1}{N}+\frac{2}{N}} )
 \cdots \Gamma({\scriptstyle n-\frac{1}{N}+\frac{N-1}{N}} )
     }
    {
    \frac{(N-1)^{(N-1) n}}
       { (2 \pi)^{\frac{N-2}{2}} \sqrt{N-1} }
 \Gamma({\scriptstyle n })\Gamma({\scriptstyle n + \frac{1}{N-1}} )
      \Gamma({\scriptstyle n+ \frac{2}{N-1}} )
 \cdots \Gamma({\scriptstyle n+ \frac{N-2}{N-1}} ) }
\\
&=&
- \frac{1}{\sqrt{2\pi} N} \sqrt{\frac{N-1}{N}}
\sum_{n=1} ^{\infty}
\frac{1 }{n!} \left( \frac{-N^{N}  e^{N(x-p)}}{(N-1)^{N-1}} \right)^n
\underbrace{
 \frac{
  \Gamma({\scriptstyle n-\frac{1}{N}} )
  \Gamma({\scriptstyle n-\frac{1}{N}+\frac{2}{N}} )
 \cdots \Gamma({\scriptstyle n-\frac{1}{N}+\frac{N-1}{N}} )
     }
    {
\Gamma({\scriptstyle n + \frac{1}{N-1} })
      \Gamma({\scriptstyle n+ \frac{2}{N-1} })
 \cdots \Gamma( {\scriptstyle n+ \frac{N-2}{N-1}} ) }
 }_{f(n,N)}
\\
&=&
- \frac{1}{\sqrt{2\pi} N} \sqrt{\frac{N-1}{N}} f(0,N)
 \left[
 \sum_{n=0} ^{\infty}
 \frac{1}{n!} \left( \frac{-N^{N}  e^{N(x-p)}}{(N-1)^{N-1}} \right)^n
 \frac{f(n,N)}{f(0,N)}
 -1
 \right] \qquad (N>1)
\\
&=&
\left( 1-\frac{1}{N} \right)
\left[
_{N -1} F _{N-2}({\scriptstyle \frac{1}{N},\frac{1}{N},\frac{2}{N},\cdots
\frac{N-2}{N}
;
\frac{1}{N-1},\frac{2}{N-1},\cdots
\frac{N-2}{N-1}; \frac{-N^{\scriptscriptstyle N} 
 e^{\scriptscriptstyle N(x-p)}}
{(N-1)^{\scriptscriptstyle N-1}}
})
-1\right].
\end{eqnarray*}
The solution of $y$ is 
\begin{eqnarray}
\bar{y}
&=&
\left( 1-\frac{1}{N} \right)
\left[
_{N -1} F _{N-2}({\scriptstyle \frac{1}{N},\frac{1}{N},\frac{2}{N},\cdots
\frac{N-2}{N}
;
\frac{1}{N-1},\frac{2}{N-1},\cdots
\frac{N-2}{N-1}; \frac{-N^{\scriptscriptstyle N} 
 e^{\scriptscriptstyle N(x-p)}}
{(N-1)^{\scriptscriptstyle N-1}}
})
-1\right]
\end{eqnarray}
where $N>1$.
In Fermion case ($N=1$),
we return to the first expression (\ref{eqn:ysolution2})
and can get the Fermion distribution function.
The hypergeometric function is represented by multiple integral,
so by analytical continuation the present solution is valid
in all temperature region.
This solution perfectly corresponds to
$g=\frac{1}{2} \quad (N=2)$ and $g=\frac{1}{3} \quad (N=3)$
which is obtained exactly in Eq. (\ref{eqn:y}).
We numerically calculate the low temperature behavior of the
3-d specific heat by using this solution.
{\flushleft {\bf Appendix C: 
Duality for the coefficient of the Sommerfeld expansion}  }

We improve the Sommerfeld expansion
and show the duality relation of the coefficient for $g$-on in this Appendix
\cite{H. Suzuki}.
The Sommerfeld expansion is the expansion in terms of small deviations from
the step function
\begin{eqnarray*}
  \int_0 ^\infty \frac{x^{s-1}}{y+g} dx
&=&
  \int_0 ^{p} x^{s-1}
  \left( \frac{1}{y+g}-\frac{1}{g} \right) dx
  +
  \int_{p} ^\infty \frac{x^{s-1}}{y+g} dx
  +
  \int_0 ^{p} \frac{x^{s-1}}{g} dx .
\end{eqnarray*}
We set
\begin{eqnarray*}
  p = \bar{p}-(1-g) \log2 
\end{eqnarray*}
to express the upper and lower value of the integral of $y$
explicitly.
Then the expansion is as follows;
\begin{eqnarray*}
  \int_0 ^\infty \frac{x^{s-1}}{y+g} dx
&=&
  \int_0 ^{\bar{p}} x^{s-1}
  \left( \frac{1}{y+g}-\frac{1}{g} \right) dx
  +
  \int_{\bar{p}} ^\infty \frac{x^{s-1}}{y+g} dx
  +
  \int_0 ^{\bar{p}} \frac{x^{s-1}}{g} dx
\\
&=&
 - \int_0 ^{\bar{p}} x^{s-1}
  \frac{y}{g(y+g)} dx
  +
  \int_{\bar{p}} ^\infty \frac{x^{s-1}}{y+g} dx
  +
  \frac{1}{g}\frac{\bar{p}^s}{s}
\\
&=&
  \int^0 _{\bar{p}} (\bar{p}-u)^{s-1}
  \frac{y}{g(y+g)} du
  +
  \int_0 ^\infty (\bar{p}+v)^{s-1} \frac{1}{y+g} dx
  +
  \frac{1}{g}\frac{\bar{p}^s}{s}.
\end{eqnarray*}
In the limit of $t \rightarrow 0$ ($\bar{p}\rightarrow \infty$),
we can approximate this integral as follows
\begin{eqnarray*}
  \int_0 ^\infty \frac{x^{s-1}}{y+g} dx
&\simeq&
 - \int_0 ^\infty (\bar{p}-u)^{s-1}
  \frac{y}{g(y+g)} du
  +
  \int_0 ^\infty (\bar{p}+v)^{s-1} \frac{1}{y+g} dx
  +
  \frac{1}{g}\frac{\bar{p}^s}{s}
\\
&=&
 - \sum_{n=0}^{\infty} \frac{ \Gamma({\scriptstyle s})
   }{\Gamma({\scriptstyle s-n}) \Gamma({\scriptstyle n+1})}
 \left( -1 \right)^n \frac{ \bar{p} ^{s-n-1}}{g}
 \int_0 ^\infty u^{n} \frac{y}{y+g} du
\\
& &
 +
 \sum_{n=0}^{\infty} \frac{ \Gamma({\scriptstyle s})
   }{\Gamma({\scriptstyle s-n}) \Gamma({\scriptstyle n+1})}
 \bar{p}^{s-n-1}  \int_0 ^\infty v^{n} \frac{1}{y+g} dv
  +
  \frac{1}{g}\frac{\bar{p}^s}{s}
\\
&=&
 - \sum_{n=0}^{\infty} \frac{ \Gamma({\scriptstyle s})
   }{\Gamma({\scriptstyle s-n}) \Gamma({\scriptstyle n+1})}
  \bar{p} ^{s-n-1} 
  \left(
 \frac{ (-1)^n }{g}
 B_n (g)
 +
 C_n (g)
   \right)
  +
  \frac{1}{g}\frac{\bar{p}^s}{s} 
\end{eqnarray*}
where $-x+\bar{p}\equiv u$, $x-\bar{p}\equiv v$ in third line,
and $B_n(g)$ and $C_n(g)$ are given by converting $u$, $v$ into $y$
\begin{eqnarray*}
  B_n({\scriptstyle g}) &\equiv&  \int_0 ^\infty u^{n} \frac{y}{y+g} du 
 =
  \int_1 ^0 \frac{dy}{y+1} 
  \left( -g \log{y} - (1-g) \log{(y+1)} + (1-g)\log2 \right)^n,
\\
  C_n({\scriptstyle g}) &\equiv&   \int_0 ^\infty v^{n} \frac{1}{y+g} dv
 =
  \int_1 ^\infty \frac{dy}{y(y+1)} 
  \left( g \log{y} + (1-g) \log{(y+1)} - (1-g)\log2 \right)^n .
\end{eqnarray*}
In $B_n({\scriptstyle g})$ and $C_n({\scriptstyle g})$,
we convert $y$ into $z=\frac{y}{y+1}$ and
obtain
\begin{eqnarray*}
  B_n({\scriptstyle g})
 &=&
  \int_{1/2} ^{0} \frac{dz}{1-z} 
  \left( -g \log{z} + \log{(1-z)} + (1-g)\log2 \right)^n,
 \\
  C_n({\scriptstyle g})
 &=&
  \int_{1/2} ^1 \frac{dz}{z} 
  \left( g \log{z} - \log{(1-z)} - (1-g)\log2 \right)^n .
\end{eqnarray*}
Moreover,
in $C_n({\scriptstyle g})$ and $B_n({\scriptstyle g})$,
we convert $z$ into $z=1-w$ and $g$ into $\frac{1}{g}$ respectively,
then
\begin{eqnarray*}
  C_n({\scriptstyle g}) 
 &=&
  - \int_{1/2} ^0 \frac{dw}{1-w} 
  \left( g \log{(1-w)} - \log{w} - (1-g)\log2 \right)^n,
\\
  B_n({\scriptstyle g}) 
 &=&
   \frac{1}{g^n} \int_{1/2} ^{0} \frac{dz}{1-z} 
  \left( - \log{z} + g \log{(1-z)} - (1-g)\log2 \right)^n .
\end{eqnarray*}
So we get the duality relation
\begin{eqnarray}
  C_n({\scriptstyle g}) 
&=&
 -g^n B_n({\scriptstyle \frac{1}{g}}) .
\end{eqnarray}
Finally,
the Sommerfeld expansion is given by
\begin{eqnarray}
   \int_0 ^\infty \frac{x^{s-1}}{y+g} dx
=
 - \sum_{n=0}^{\infty} \frac{ \Gamma({\scriptstyle s})
   }{\Gamma({\scriptstyle s-n}) \Gamma({\scriptstyle n+1})}
  \bar{p} ^{s-n-1} 
  \left(
 \frac{ (-1)^n }{g}
 B_n (g)
 -g^n B_n({\scriptstyle \frac{1}{g}})
   \right)
  +
  \frac{1}{g}\frac{\bar{p}^s}{s} .
\end{eqnarray}
The Fermion case ($g=1$),
the odd number of $n$ in the sum does not cancel.

\newpage

\end{document}